\documentclass[a4paper,11pt]{article}
\usepackage{jinstpub} 

\usepackage{amssymb}
\usepackage{amsmath}
\usepackage{graphicx}
\usepackage[colorlinks=true,linkcolor=blue,citecolor=blue,urlcolor=blue]{hyperref}
\usepackage{siunitx}
\usepackage{booktabs}
\usepackage{multirow}
\usepackage{xcolor}

\title{\boldmath Simulation study of accelerator-based muography\\using the GeV-scale forward muon component at SHINE}

\author[a]{Jiangtao~Wang,}
\author[a]{Yinghe~Wang,}
\author[b]{Si Chen,}
\author[b]{Rongbing Deng,}
\author[a,1]{Kim~Siang~Khaw,\note{Corresponding author.}}
\emailAdd{kimsiang84@sjtu.edu.cn}
\author[a]{Jun~Kai~Ng,}
\author[b]{Qisheng Tang,}
\author[c]{Dong Wang,}
\author[b]{Wenzhen Xu,}
\author[a]{and Jia~Cheng~Yap}

\affiliation[a]{State Key Laboratory of Dark Matter Physics, Key Laboratory for Particle Astrophysics and Cosmology (MOE), Shanghai Key Laboratory for Particle Physics and Cosmology (SKLPPC),  Tsung-Dao Lee Institute \& School of Physics and Astronomy, Shanghai Jiao Tong University, \\Shanghai 201210, China}
\affiliation[b]{Shanghai Advanced Research Institute, Chinese Academy of Sciences, \\Shanghai 201210, China}
\affiliation[c]{Center for Transformative Science (CTS), ShanghaiTech University, \\Shanghai 201210, China}

\abstract{
Muography exploits the penetrating power of muons to image the interior of large dense objects, but cosmic-ray sources deliver only $\sim 1~\text{cm}^{-2}\,\text{min}^{-1}$ predominantly from above, limiting both imaging speed and accessible geometries.
Electron-driven muon production has recently been demonstrated with laser-wakefield accelerators, yet their shot-to-shot fluctuations hinder the systematic studies required for quantitative accelerator muography.
Using Geant4 Monte Carlo simulations, we model the full experimental setup at Shaft~2 of the Shanghai High repetition rate XFEL and Extreme light facility (SHINE), from the interaction of a \SI{3}{GeV}, \SI{50}{pC}, \SI{50}{Hz} commissioning electron beam with the muon target through \SI{25}{m} of beamline structures, including a \SI{3}{m}-thick concrete isolation wall.
Approximately 0.28 effective reconstructed single-muon events per bunch, with residual kinetic energies below \SI{1.2}{GeV} after traversing the wall, reach the downstream muography test area (a rate of $\sim$\SI{14}{\per\second}), while the isolation wall absorbs most charged background particles with kinetic energies below $\sim$\SI{1}{GeV}; the residual neutron background can be discriminated using its characteristic energy deposition in the detector.
Scattering-tomography simulations show that $3\times10^5$ effective muon events (accumulated in approximately \SI{6}{h} at the commissioning rate) yield a Structural Similarity Index above 0.9, demonstrating the feasibility of quantitative accelerator-based muography with the SHINE electron-driven GeV-scale forward muon source.
At the \SI{8}{GeV}/\SI{50}{kHz} muography benchmark, the projected available muon intensity exceeds $5\times10^{4}~\mu/\text{s}$, corresponding to a conservative low-occupancy operating scale of approximately one muon arriving at the detector per bunch. This intensity exceeds the cosmic-ray flux by orders of magnitude, while the stability of SHINE's superconducting linac makes the facility a controlled platform for developing the electron-on-target muography technique.
}

\arxivnumber{} 

\begin{document}
\maketitle
\section{Introduction}
\label{sec:intro}

Muon imaging, or muography, uses the penetrating power of muons to probe the internal density structure of large and otherwise inaccessible objects~\cite{Bonechi2020, Procureur2018}. By measuring either the attenuation of muons or their deflection by multiple Coulomb scattering, it is possible to reconstruct projected density distributions and, with suitable geometries, three-dimensional material structures. Since the pioneering work of Alvarez~et~al.~\cite{Alvarez1970}, muography has developed into a versatile tool for volcanology~\cite{Tanaka2007}, civil engineering~\cite{Guardincerri2017}, nuclear-waste monitoring~\cite{Clarkson2015}, and cultural-heritage studies~\cite{Morishima2017,ScanPyramids2023}.

The principal limitation of conventional muography is the source itself. Cosmic-ray muons arrive at sea level with a mean energy of approximately \SI{4}{GeV}, an integrated vertical flux of about $1~\text{cm}^{-2}\,\text{min}^{-1}$, and an angular distribution that is approximately proportional to $\cos^2\theta$~\cite{PDG2024}. Their low flux can require exposure times ranging from hours to months, while their predominantly downward direction restricts the geometries that can be explored. An accelerator-based muon source could address both limitations by providing a controllable beam direction, a well-defined time structure, and a substantially higher usable rate.

Electron-driven muon production has recently been demonstrated with laser-wakefield-accelerated (LWFA) electron beams, including experiments at the SULF \SI{1}{PW} laser~\cite{Zhang2025NatPhys}, BELLA~\cite{Terzani2025PRAB}, and ELI-NP~\cite{Calvin2026PPCF}. These experiments have established the basic feasibility of compact electron-on-target muon production. For quantitative muography, however, systematic studies also require reproducible beam conditions. Present LWFA sources can exhibit substantial shot-to-shot variations in electron energy, charge, and pointing, making it difficult to disentangle source fluctuations from detector and reconstruction performance.

In this work, we investigate a complementary approach based on the Shanghai High-repetition-rate XFEL and Extreme light facility (SHINE)~\cite{Zhao2018SHINE}. We ask a simple practical question: can the GeV-scale forward muon component produced by electron-on-target interactions at SHINE be used directly for scattering-based muography? Previous studies of the SHINE muon program have focused on the low-energy surface-muon component for $\mu$SR and precision muon physics~\cite{Lv2023IPAC,Liu2025PRAB}. The same target, however, also produces a higher-energy and more forward-directed muon component. Here we follow that component through the existing Shaft~2 infrastructure and evaluate whether the surviving muons can support a realistic imaging measurement.

The central idea of this study is that the existing Shaft~2 shielding is not only an engineering constraint but also a useful part of the source system. The shielding suppresses the intense electromagnetic and hadronic shower produced at the target while preferentially transmitting the more penetrating GeV-scale muons. We use Geant4 simulations to model the full configuration, from a \SI{3}{GeV}, \SI{50}{pC}, \SI{50}{Hz} commissioning electron beam incident on the target to the downstream imaging area behind a \SI{3}{m}-thick concrete isolation wall. We then use the simulated muon phase space as input to a scattering-tomography simulation.

The resulting picture is encouraging. Approximately 0.28 effective reconstructed single-muon events per bunch are obtained for the simulated detector configuration, corresponding to about \SI{14}{\per\second} at \SI{50}{Hz}. The muons that reach the test area have residual kinetic energies up to \SI{1.2}{GeV}, while most charged backgrounds have been removed by the isolation wall. In the imaging simulation, $3\times10^5$ effective muon events are sufficient to reach an SSIM above 0.9, corresponding to approximately \SI{6}{h} at the commissioning rate. These results suggest that SHINE can serve not only as a high-intensity muon source, but also as a controlled test bed for developing accelerator-based muography.

The remainder of the article follows this physical chain. Section~\ref{sec:production} identifies the muon component relevant for muography. Section~\ref{sec:facility} introduces the Shaft~2 geometry and its role as a passive particle filter. Section~\ref{sec:simulation} characterizes the particle environment and the usable muon beam behind the wall. Section~\ref{sec:imaging} evaluates the corresponding imaging performance, and Section~\ref{sec:discussion} discusses the broader implications, limitations, and next experimental steps.

\section{GeV-scale forward muon production at SHINE}
\label{sec:production}

\begin{figure}[htbp]
\centering
\includegraphics[width=0.7\linewidth]{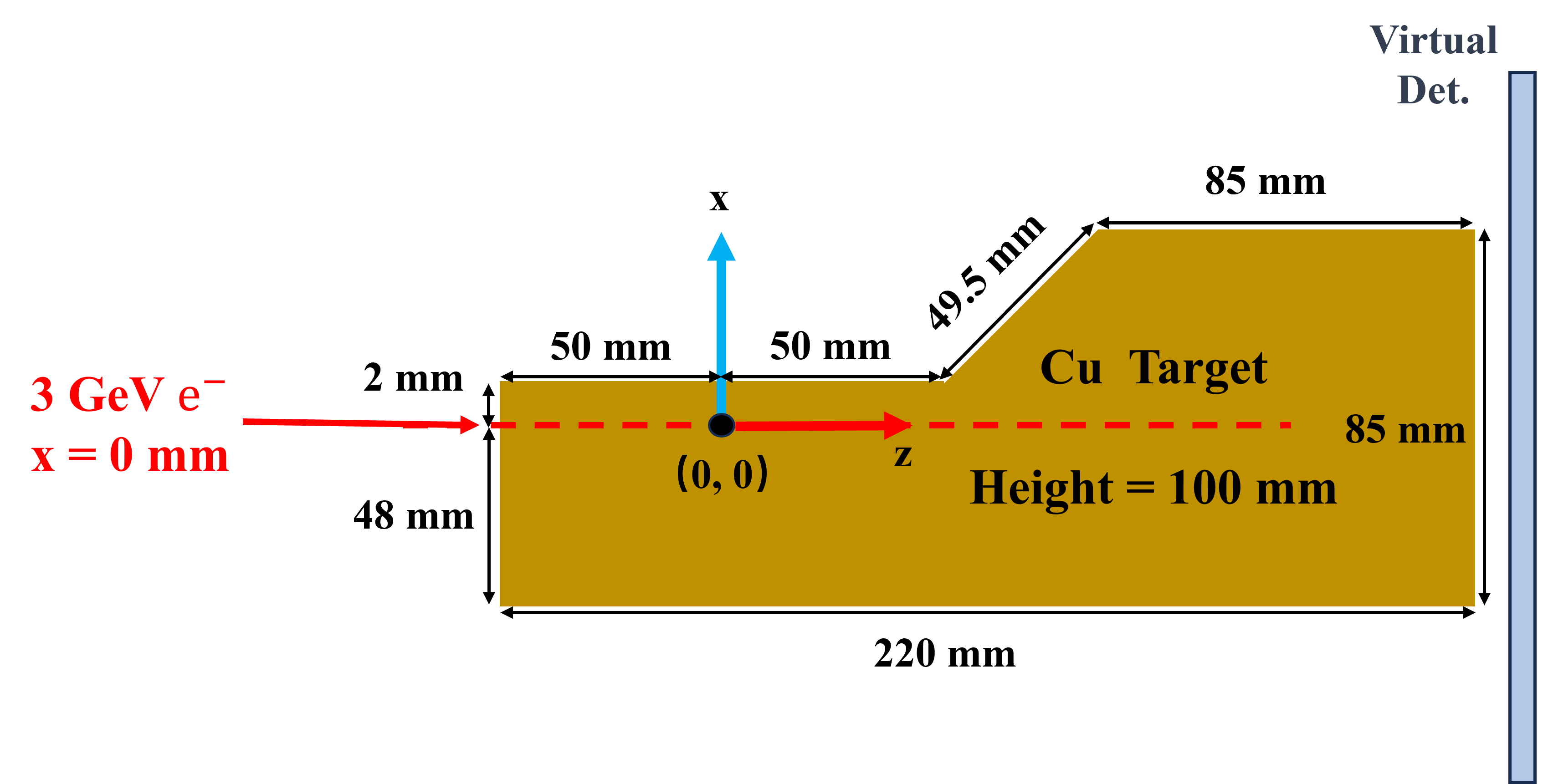}
\caption{Cross-sectional view of the muon target in the X-Z plane. The target material is copper. The structure is symmetric about the y-axis.}
\label{fig:MuonTarget}
\end{figure}

The muon target is shown in Fig.~\ref{fig:MuonTarget}. The \SI{3}{GeV} electron beam is incident along the $z$-axis, and a \SI{1}{m} $\times$ \SI{1}{m} virtual detector is placed downstream of the target to record the emerging secondary particles. Two electron bunches, corresponding to a total charge of \SI{100}{pC}, were simulated for the production-mechanism study.

When the electron beam interacts with the thick copper target, muons are produced through several processes. For the present application, two channels are particularly important: photonuclear pion production and Bethe--Heitler muon-pair production. Although both contribute to the total muon yield, they populate very different regions of phase space. This distinction is central to the present work because the existing shielding strongly favors the more energetic and forward-directed component.

In the photonuclear channel, bremsstrahlung photons generated in the electron shower excite the target nuclei and produce pions, which subsequently decay through $\pi^\pm \to \mu^\pm + \nu_\mu/\bar{\nu}_\mu$~\cite{Zhang2025NatPhys,Terzani2025PRAB}. In the inclusive simulation, muons recorded by the downstream virtual detector were classified according to their production process. The photonuclear contribution yields $9\,989\pm71$ muons per electron bunch and dominates the total yield. Most of these muons have kinetic energies below approximately \SI{1}{GeV} and occupy a relatively broad angular distribution.

The second component arises from Bethe--Heitler pair production, in which bremsstrahlung photons convert into $\mu^+\mu^-$ pairs in the Coulomb field of a nucleus,
\begin{equation}
e^- + Z_1 \to Z_1 + \gamma, \qquad \gamma + Z_2 \to \mu^+ + \mu^- + Z_2.
\end{equation}
The cross section is roughly three orders of magnitude smaller than that of the dominant photonuclear channel~\cite{Nagamine2009,Blomqvist1977}, but the resulting muons are typically more energetic and more strongly concentrated around the beam direction~\cite{Lv2023IPAC,Liu2025PRAB}. The mutually exclusive Bethe--Heitler contribution yields $1\,276\pm25$ muons per bunch in the same scoring plane. Muons can also be produced through the trident process, $e^-+Z\to e^-+Z+\mu^++\mu^-$, although Bethe--Heitler production dominates the direct pair-production contribution in a thick target~\cite{Calvin2026PPCF}.

\begin{figure}[htbp]
\centering
\includegraphics[width=1.0\linewidth]{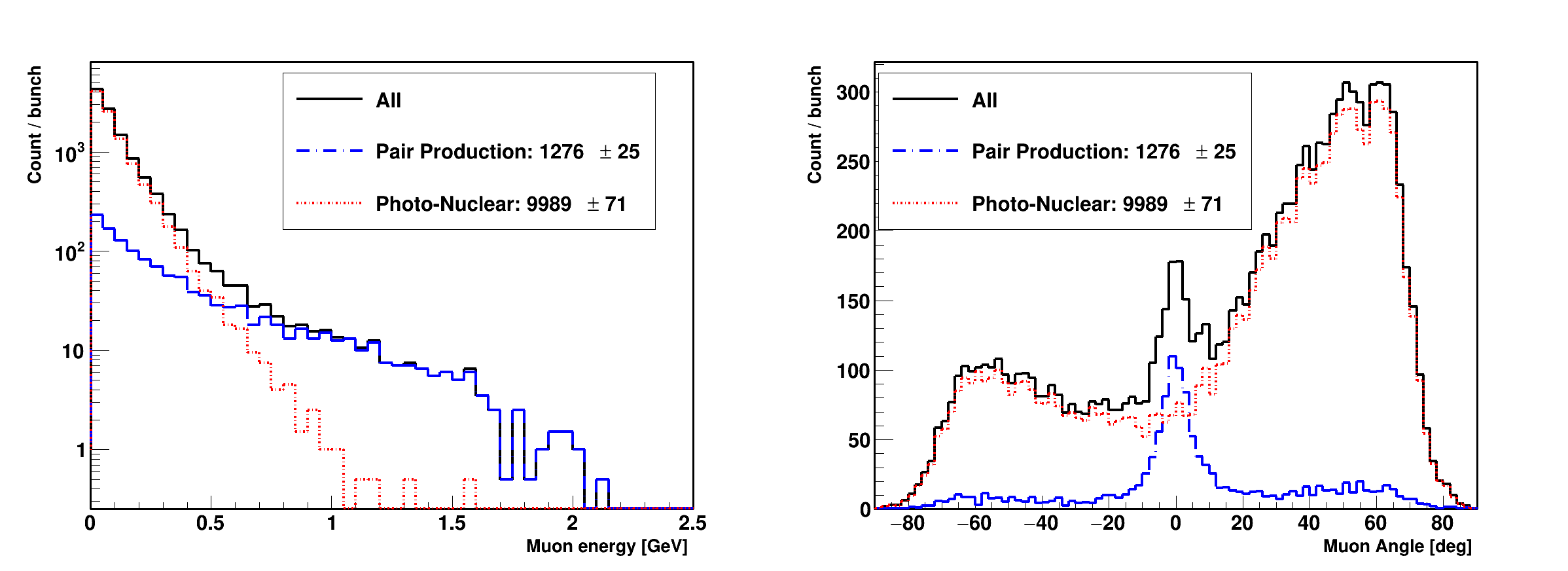}
\caption{Simulated muon distributions at the target exit for a \SI{3}{GeV}, \SI{50}{pC}/bunch electron beam, decomposed by production mechanism. (\textbf{a}) Kinetic energy spectrum: the photonuclear channel dominates the lower-energy region ($E_k \lesssim \SI{1}{GeV}$), whereas Bethe--Heitler pair production generates a higher-energy component. (\textbf{b}) Angular distribution of muon momenta in the $xz$ plane, illustrating the more forward-peaked nature of the Bethe--Heitler component.}
\label{fig:production_spectrum}
\end{figure}

Figure~\ref{fig:production_spectrum} therefore already suggests the component that is most relevant for muography. Photonuclear production provides the larger overall yield, whereas the Bethe--Heitler contribution supplies the GeV-scale forward muons most likely to penetrate the downstream shielding. Rather than attempting to transport the entire secondary distribution, the Shaft~2 geometry naturally selects this penetrating part of phase space. The next section describes how the existing facility layout performs that selection.

\section{SHINE Shaft~2 as a natural muon filter}
\label{sec:facility}

\subsection{SHINE facility and beam parameters}
\label{sec:facility:beam}

SHINE~\cite{Zhao2018SHINE} is a fourth-generation light source under construction in Zhangjiang, Shanghai.
Its \SI{8}{GeV} CW superconducting RF linac is designed to deliver a bunched electron beam at up to \SI{1}{MHz} with \SI{100}{pC}/bunch (\SI{100}{\micro\ampere} average current); muon-source operation at the dedicated target is planned at \SI{50}{kHz}~\cite{Liu2025PRAB}.
Three undulator beamlines produce hard X-rays up to \SI{25}{keV}; after lasing, the electron beams are diverted to beam dumps.
The SHINE beam dump (cylindrical Al/Cu composite, \SI{16}{cm} radius, \SI{115}{cm} length) produces muon yields of $\sim 5 \times 10^3~\mu^+$/bunch at \SI{8}{GeV} in simulation~\cite{Lv2023IPAC}.
A dedicated copper target, optimized for surface muon production, can deliver $\sim 2 \times 10^3$ surface muons per bunch, corresponding to $3 \times 10^6~\mu^+$/s at \SI{50}{kHz}~\cite{Liu2025PRAB}.

During the SHINE commissioning phase, the \SI{3}{GeV} configuration with \SI{50}{pC} at \SI{50}{Hz} is used for the muography studies. Throughout this article, one \emph{bunch} denotes a single \SI{50}{pC} bunch delivered in one \SI{50}{Hz} cycle. 

\subsection{Shaft~2 layout}
\label{sec:facility:layout}

\begin{figure}[htbp]
\centering
\includegraphics[width=0.9\linewidth]{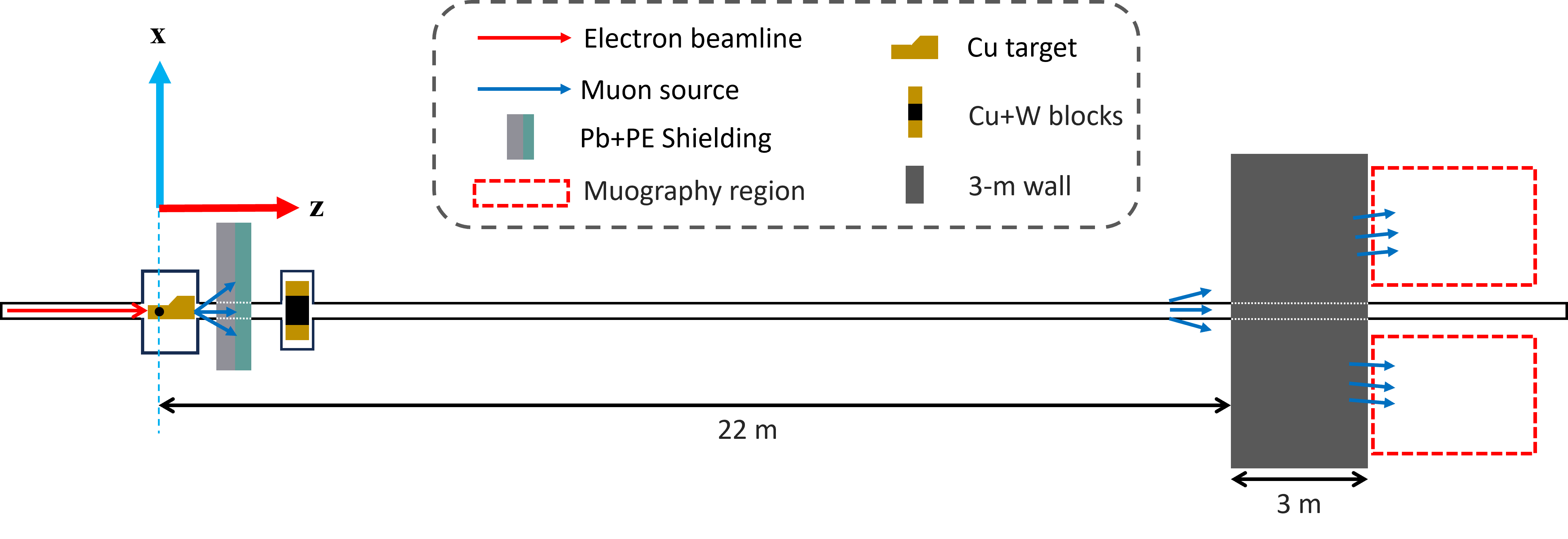}
\caption{
Schematic layout of the accelerator muography experiment at SHINE Shaft~2. The coordinate system is defined with $z$ along the beam axis and $x$ in the transverse horizontal direction. The electron beam enters from the left and interacts with the target. The key structures along the beam axis ($z$) are labeled with their longitudinal positions. High-energy muons penetrate the \SI{3}{m}-thick isolation wall and reach the downstream test area.
}
\label{fig:layout}
\end{figure}

\begin{figure}[htbp]
\centering
\includegraphics[width=0.8\linewidth]{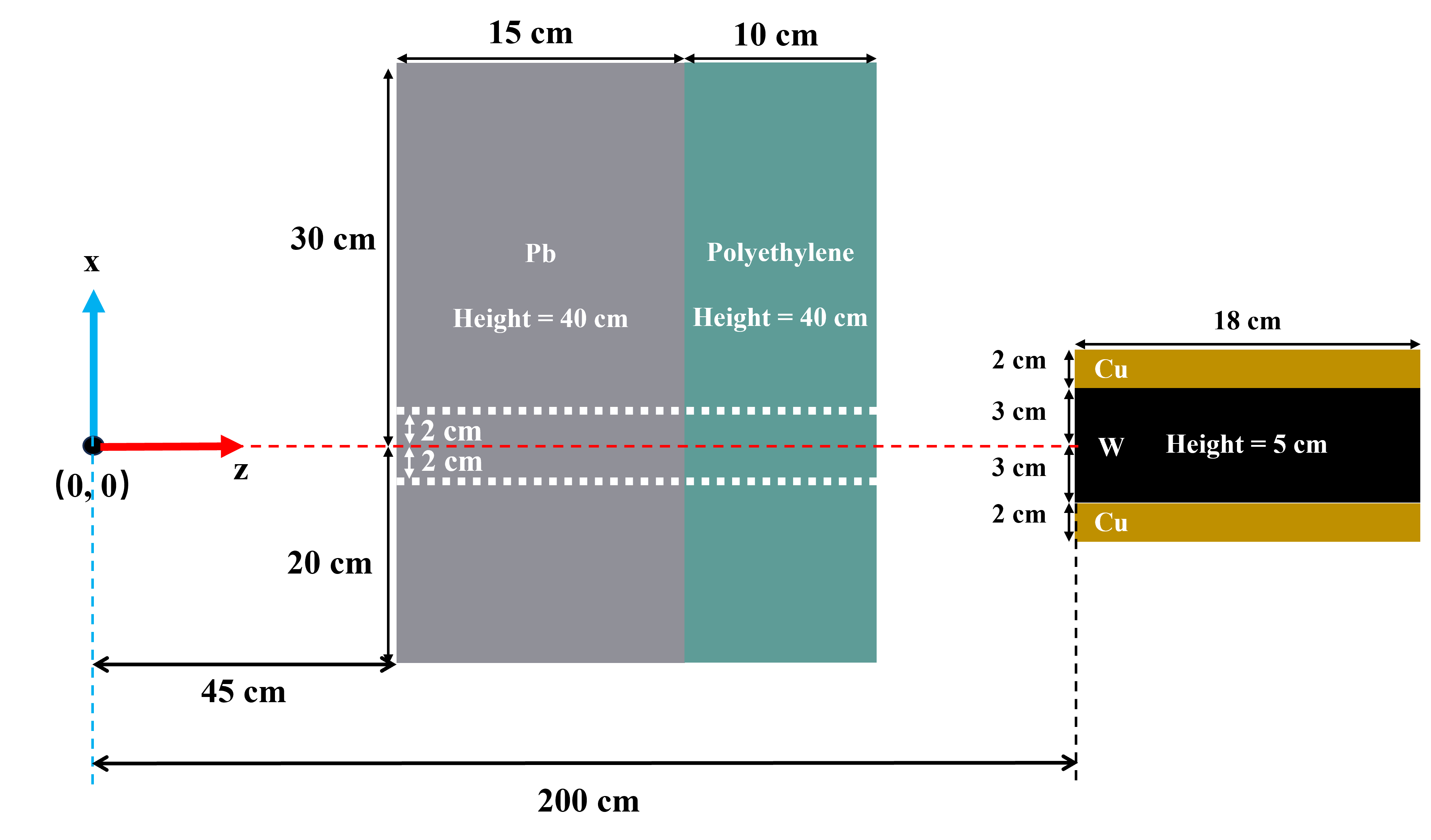}
\caption{Cross-sectional view of the shielding blocks upstream of the isolation wall. From upstream to downstream: a lead (Pb) block (\SIrange{0.45}{0.6}{m}), a polyethylene (PE) block (\SIrange{0.6}{0.7}{m}), and copper--tungsten (Cu, W) blocks (\SIrange{2.0}{2.18}{m}). A 4~cm-diameter vacuum pipe passes through these structures along the beam axis ($z$). The longitudinal block positions are given in meters, while the transverse dimensions labeled in the figure are in centimeters.}
\label{fig:Shielding}
\end{figure}

The proposed muography area is located in Shaft~2, downstream of the target and behind a \SI{3}{m}-thick concrete isolation wall situated approximately \SIrange{22}{25}{m} from the production point, as shown in Fig.~\ref{fig:layout}. This geometry is particularly useful for the present study because the particles must pass through the same shielding that separates the accelerator environment from the downstream experimental area.
The beamline between the target and the test area includes several structures relevant to muon transport: a lead shielding block (spanning $z =$ \SIrange{0.45}{0.6}{m}), a polyethylene (PE) shielding block ($z =$ \SIrange{0.6}{0.7}{m}), copper--tungsten blocks ($z =$ \SIrange{2.0}{2.18}{m}), and the isolation wall itself. The detailed parameters are shown in Figs.~\ref{fig:MuonTarget} and~\ref{fig:Shielding}. In the Pb shielding, PE shielding, and isolation wall, a 4-cm-diameter vacuum pipe passes through these structures along the beam axis.

Together, these structures progressively attenuate the electromagnetic and hadronic shower. High-energy muons are much less affected because their energy loss is dominated by ionization, at approximately \SI{2}{MeV\,cm^2/g} over the relevant energy range~\cite{PDG2024}. The downstream area therefore receives a particle distribution that is already strongly filtered before any detector-based selection is applied.

\section{Monte Carlo simulation}
\label{sec:simulation}

\subsection{Simulation setup}
\label{sec:sim:setup}

The simulations were performed with the Geant4 toolkit~\cite{Agostinelli2003,Allison2016} using the \texttt{FTFP\_BERT} physics list, which combines the Fritiof string model with the Bertini cascade and pre-compound models. The geometry includes the muon target station and vacuum system, the local lead and polyethylene shielding, the copper--tungsten blocks, and the \SI{3}{m} isolation wall. In other words, the simulation follows the particles through the same structures that define the proposed experimental configuration rather than through an idealized beamline.

The coordinate system is defined with $z$ along the electron-beam axis and $x$ in the transverse horizontal direction. Virtual detector planes are placed perpendicular to the $z$-axis and have a $y$ range of \SIrange{-0.25}{0.25}{m}, comparable to the transverse size of the imaging detector. By moving these planes along the beam direction, we record the position, kinetic energy, and species of the particles that survive each stage of the infrastructure. The primary beam consists of \SI{50}{pC} of \SI{3}{GeV} electrons.

A \SI{150}{MeV} kinetic-energy cut was applied to background particles to reduce computation time. This value is close to the pion-production threshold for the photonuclear process~\cite{Dreschsel1992}, so the muon and pion production relevant to the present study are unaffected.

\subsection{Particle flux distributions}
\label{sec:sim:flux}

\begin{figure}[htbp]
\centering
\includegraphics[width=0.9\linewidth]{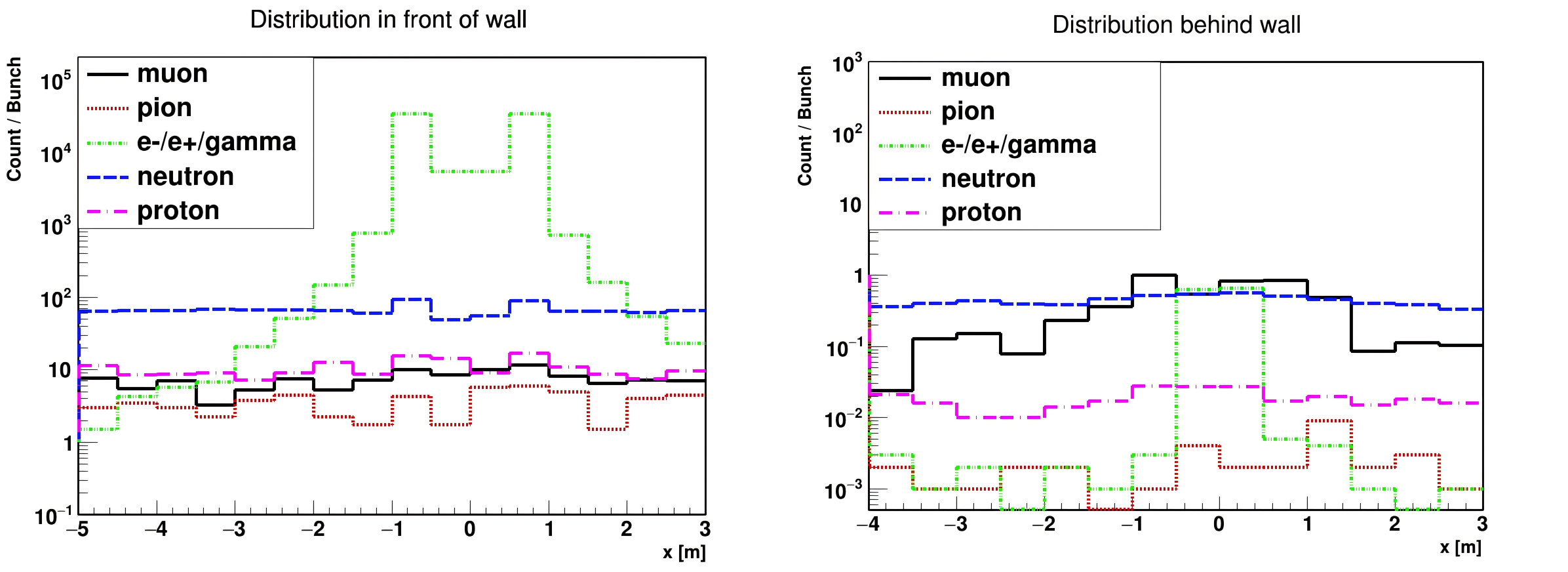}
\caption{Transverse particle flux distributions at two longitudinal positions for a \SI{3}{GeV}, \SI{50}{pC} electron beam: (\textbf{a}) $z = \SI{21}{m}$ (immediately in front of the wall), and (\textbf{b}) $z = \SI{25}{m}$ (behind the wall). Species shown: muons (black), pions (red), $e^\pm/\gamma$ (green), neutrons (blue), and protons (magenta). Background particles in front of the wall are scored with an energy threshold $E_k \ge \SI{150}{MeV}$, corresponding to the pion production threshold.}
\label{fig:flux_dist}
\end{figure}

The particle distributions were obtained using ten simulated electron bunches. Fig.~\ref{fig:flux_dist} shows the transverse particle flux distributions at selected $z$-positions for five species: muons, pions, $e^\pm/\gamma$ ($E_k \ge \SI{150}{MeV}$), neutrons ($E_k \ge \SI{150}{MeV}$), and protons ($E_k \ge \SI{150}{MeV}$).
In front of the wall ($z = \SI{21}{m}$), the flux of background particles exceeds the muon flux by 1--2 orders of magnitude, making this region unsuitable for muography. These simulated particle distributions were reused 1,000 times to improve the statistical precision of the particle distributions behind the wall.
Behind the wall ($z = \SI{25}{m}$), high-energy muons survive among the charged species, and the residual background is dominated by neutrons, which interact weakly with plastic scintillator detectors and can be further suppressed by energy deposition in detectors.

\subsection{Muon energy filtering by the isolation wall}
\label{sec:sim:wall}

\begin{figure}[htbp]
\centering
\includegraphics[width=0.6\linewidth]{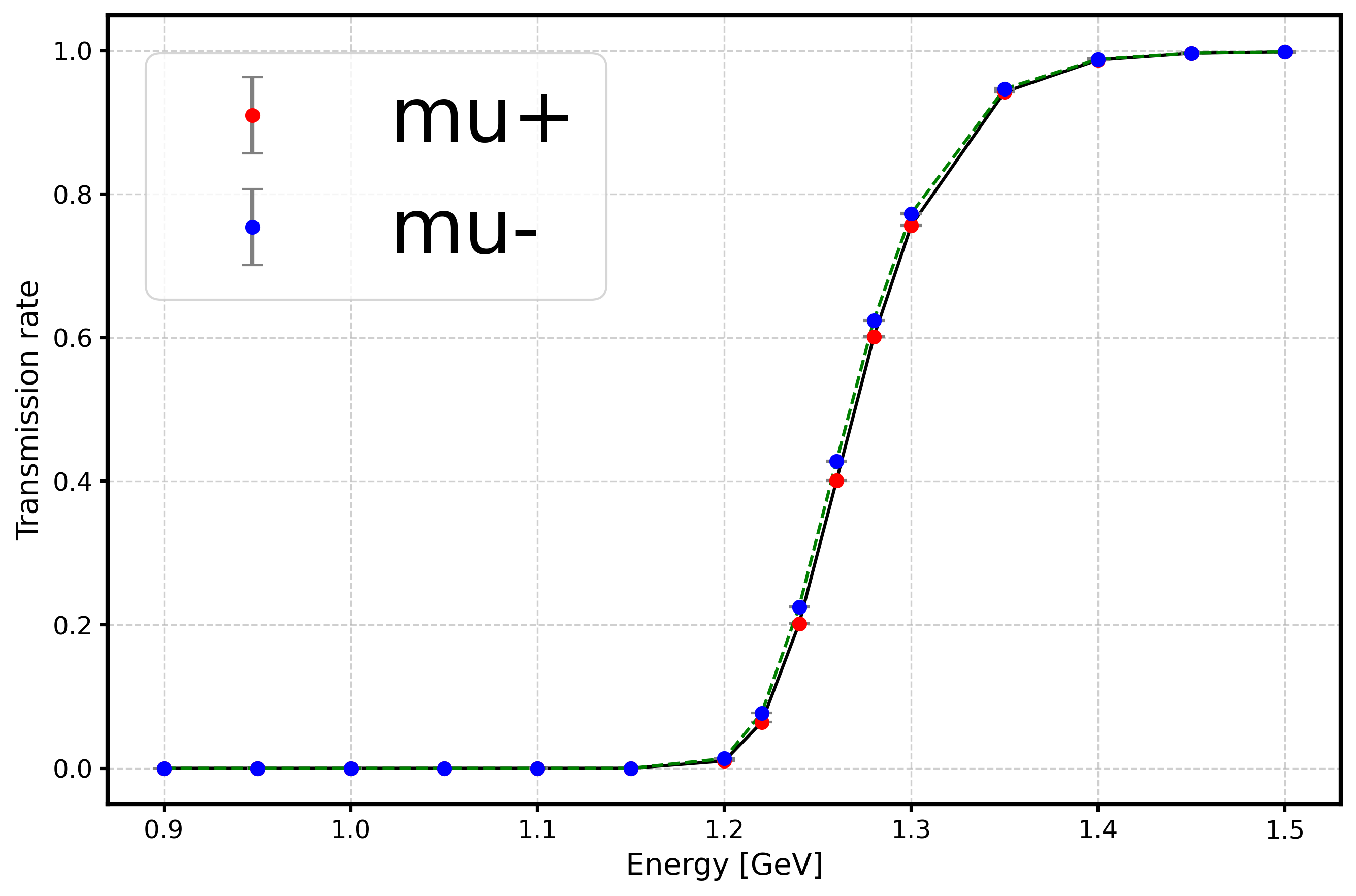}
\caption{Transmission rate of the \SI{3}{m}-thick concrete isolation wall as a function of incident muon kinetic energy, obtained from dedicated simulations with monoenergetic muons. Muons with $E_k \le \SI{1}{GeV}$ are completely absorbed. Error bars represent the statistical uncertainty from the simulation.
}
\label{fig:wall_transmission}
\end{figure}

The $\sim$\SI{1}{GeV} transmission threshold of the wall was verified with dedicated simulations in which monoenergetic muons were propagated through the wall geometry (Fig.~\ref{fig:wall_transmission}).
The threshold is consistent with the PDG data~\cite{PDG2024}. The penetration range of a 1-GeV muon in concrete is $\sim$\SI{2.4}{m}. 
Therefore, an energy threshold of \SI{1}{GeV} is adopted for the subsequent simulations of the muon distribution behind the wall. Muons above this threshold lose a large fraction of their energy while traversing the wall and emerge with degraded kinetic energies; consequently, the muon spectrum at the test area extends from \SI{0}{GeV} up to \SI{1.2}{GeV} (Section~\ref{sec:sim:rate}), even though the wall absorbs all incident muons below the threshold.

\subsection{Muon rate}
\label{sec:sim:rate}

\begin{figure}[htbp]
\centering
\includegraphics[width=0.9\linewidth]{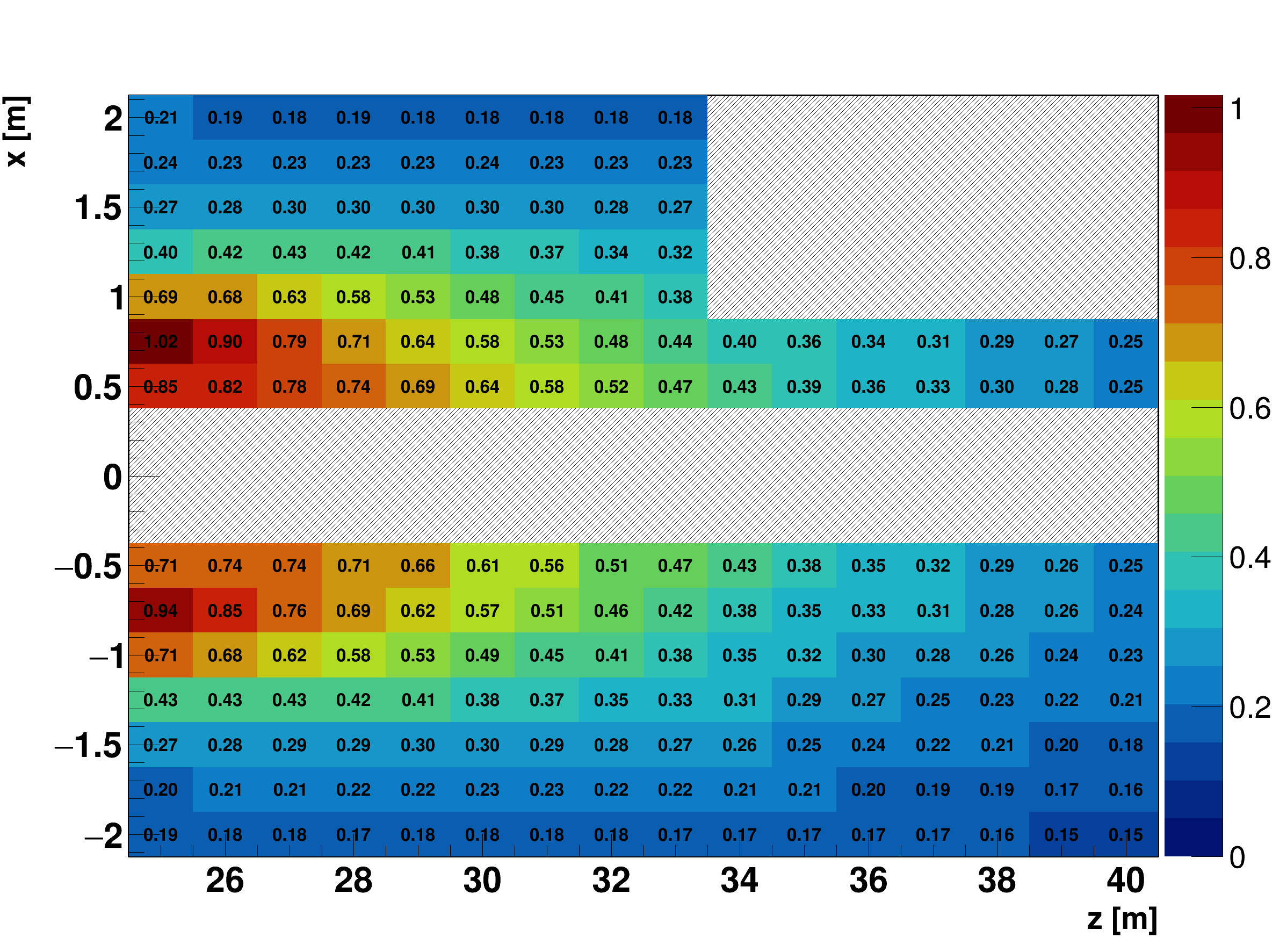}
\caption{Muon event rate as a function of detector position behind the isolation wall, simulated with $1.5\times10^5$ bunches using a $\SI{50}{cm} \times \SI{50}{cm}$ virtual detector. The color scale indicates the number of muons per bunch. Detectors cannot be placed in the shadow area owing to the presence of the vacuum pipe and shielding walls. The vertical axis ($x$) represents the transverse offset of the detector center from the beam axis, and the horizontal axis ($z$) corresponds to the position behind the wall. The maximum rate of $(1.02 \pm 0.01)$~muons/bunch is observed at $(z,\;x) = (\SI{25}{m},\; \SI{0.75}{m})$.}
\label{fig:muon_eventsRate}
\end{figure}

The primary particle distributions in front of the wall were obtained using 150 simulated electron bunches, with a \SI{1}{GeV} kinetic-energy threshold applied to background particles. The resulting particle distributions were then reused 1,000 times. The muon rate therefore corresponds to $1.5\times10^5$ equivalent electron bunches for a $\SI{50}{cm} \times \SI{50}{cm}$ virtual detector. From Fig.~\ref{fig:muon_eventsRate}, the maximum muon yield ($(1.02\pm0.01)$ muons per bunch) is observed at $z = \SI{25}{m}$ and $x = \SI{0.75}{m}$.

\begin{figure}[htbp]
\centering
\includegraphics[width=0.8\linewidth]{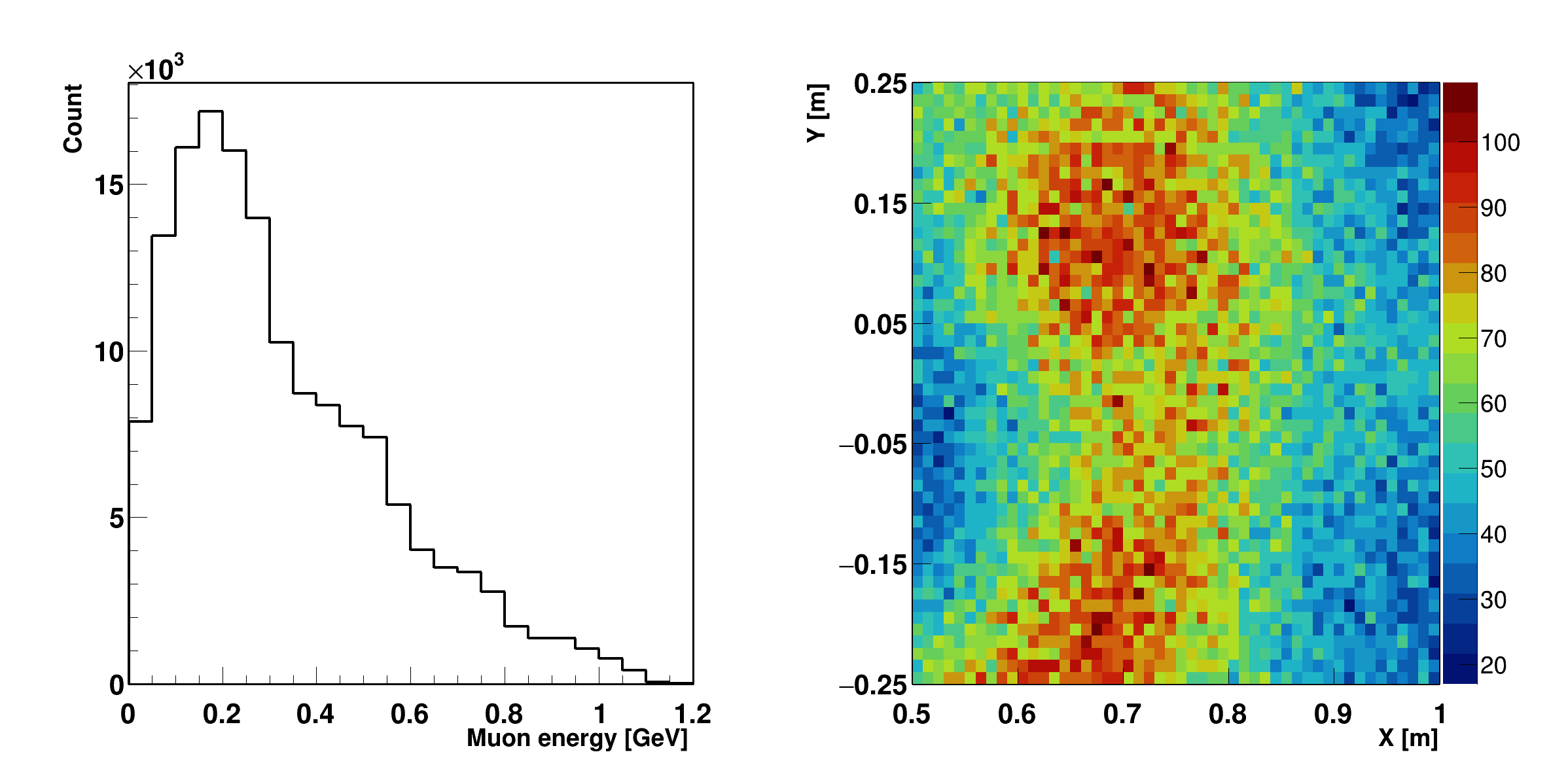}
\caption{Properties of the muon beam at the optimal detector position $(z,\;x) = (\SI{25}{m},\;\SI{0.75}{m})$ behind the isolation wall, used as input for the imaging simulations of Section~\ref{sec:imaging}. (\textbf{a}) Kinetic energy spectrum showing muons with residual energies from \SIrange{0}{1.2}{GeV} after traversing the wall. (\textbf{b}) Transverse position ($x$-projection) showing a central distribution across the detector acceptance.}
\label{fig:beam_characterization}
\end{figure}

Muon kinetic energies at the test area, after energy loss in the wall, range from \SIrange{0}{1.2}{GeV}, and the muons are centrally distributed within the detector acceptance in the $x$ direction (Fig.~\ref{fig:beam_characterization}).

Table~\ref{tab:muon_rate} summarizes the projected muon rates behind the isolation wall for the commissioning configuration and for the \SI{8}{GeV}/\SI{50}{kHz} muography benchmark.
For the \SI{3}{GeV} commissioning configuration, the simulated maximum detector-arrival rate is $\sim$\SI{51}{\per\second}. At the \SI{8}{GeV}/\SI{50}{kHz} muography benchmark, the projected available muon intensity is $>5\times 10^4~\mu/\text{s}$, corresponding to a conservative low-occupancy operating scale of approximately one muon arriving at the detector per bunch. This projected source/detector-arrival intensity should not be interpreted as a fully simulated effective reconstructed event rate. In future work, multiple-track reconstruction may allow operation at higher detector occupancy.

\begin{table}[htbp]
\centering
\caption{Muon intensity benchmarks at the muography area. The \SI{3}{GeV} entry is a simulated detector-arrival rate, whereas the \SI{8}{GeV}/\SI{50}{kHz} entry is a projected low-occupancy intensity benchmark rather than a fully simulated effective reconstructed event rate.
}
\label{tab:muon_rate}
\begin{tabular}{cccccc}
\toprule
Beam energy & Bunch charge & Rep.\ rate & $\mu$/bunch & $\mu$ rate & Remarks \\
{[\si{GeV}]} & {[\si{pC}]} & {[\si{Hz}]} &  & {[\si{\per\second}]} & \\
\midrule
3 & 50 & 50 & $\sim$1 & $\sim$50 & Commissioning\\
8 & 100 & 50\,000 & $\sim$1 arriving at detector & $> 5\times10^4$  & Muography benchmark\\
\bottomrule
\end{tabular}
\end{table}

\section{Simulated imaging performance}
\label{sec:imaging}

\subsection{Imaging simulation setup}
\label{sec:imaging:setup}

\begin{figure}[htbp]
\centering
\includegraphics[width=0.9\linewidth]{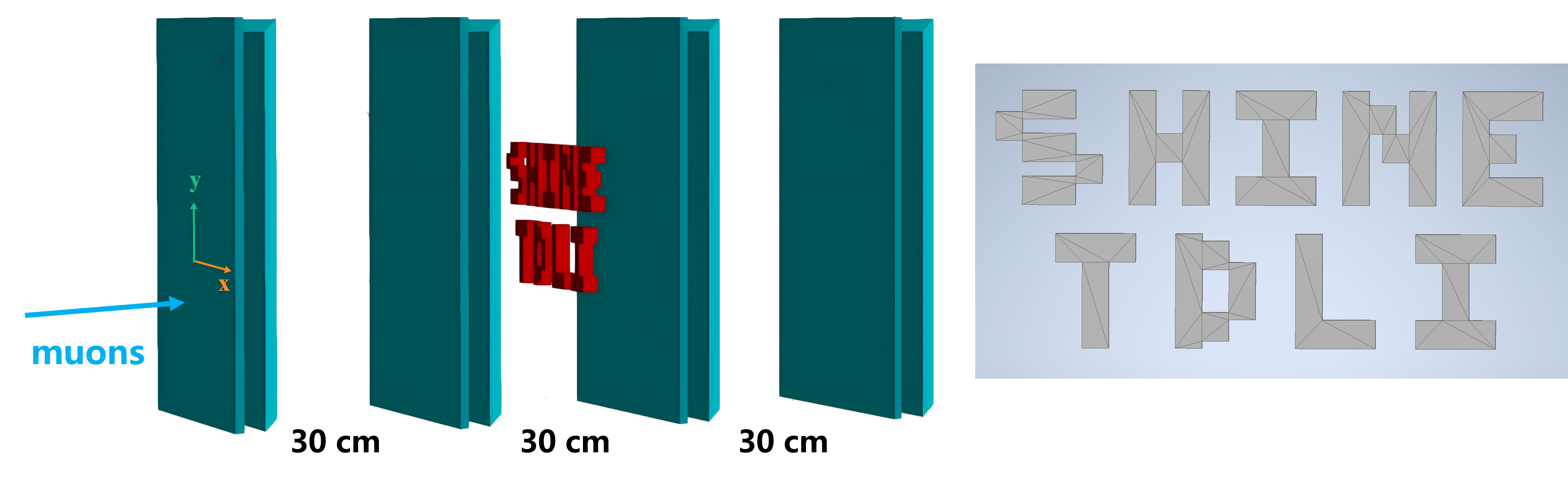}
\caption{Geometry of the muography imaging simulation at the Shaft~2 test area. The detector system consists of four tracking layers (T1--T4), each comprising two orthogonally oriented sublayers of 32 triangular-prism plastic scintillators, enabling three-dimensional coordinate reconstruction. The inter-layer spacing is \SI{30}{cm}. The test phantom (right) consists of multiple $\SI{2}{cm} \times \SI{2}{cm} \times \SI{2}{cm}$ lead cubes arranged in a predefined pattern to evaluate imaging performance.}
\label{fig:equipment}
\end{figure}

To test whether the filtered muon beam can form a useful image, the simulated phase space at the optimal Shaft~2 location was used as the source for the geometry shown in Fig.~\ref{fig:equipment}. The detector consists of four tracking layers (T1--T4). Each layer contains two orthogonally oriented sublayers of 32 triangular-prism plastic scintillators, allowing the three-dimensional trajectory of each muon to be reconstructed. Adjacent tracking layers are separated by \SI{30}{cm}. The imaging phantom is composed of lead cubes with a side length of \SI{2}{cm}.

\subsection{Scattering tomography}
\label{sec:imaging:tomography}

The trajectory of a muon is deflected by Coulomb scattering when it traverses an object~\cite{Lynch1991}.
The deflection angle follows a Gaussian distribution, and $\theta_{rms}$ is the root-mean-square value of the muon scattering angle. The approximate expression for $\theta_{rms}$ is given by:

\begin{equation}
    \theta_{rms}=\frac{13.6}{\beta c p \ [\mathrm{MeV}]} \sqrt{\frac{L}{L_0}}\left[1+0.038 \ln \left(\frac{L}{L_0}\right)\right]
\end{equation}

where $\beta c$ is the muon velocity, $p$ is the muon momentum, $L$ is the path length through the object, and $L_0$ is its radiation length.

\begin{figure}[htbp]
\centering
\includegraphics[width=0.9\linewidth]{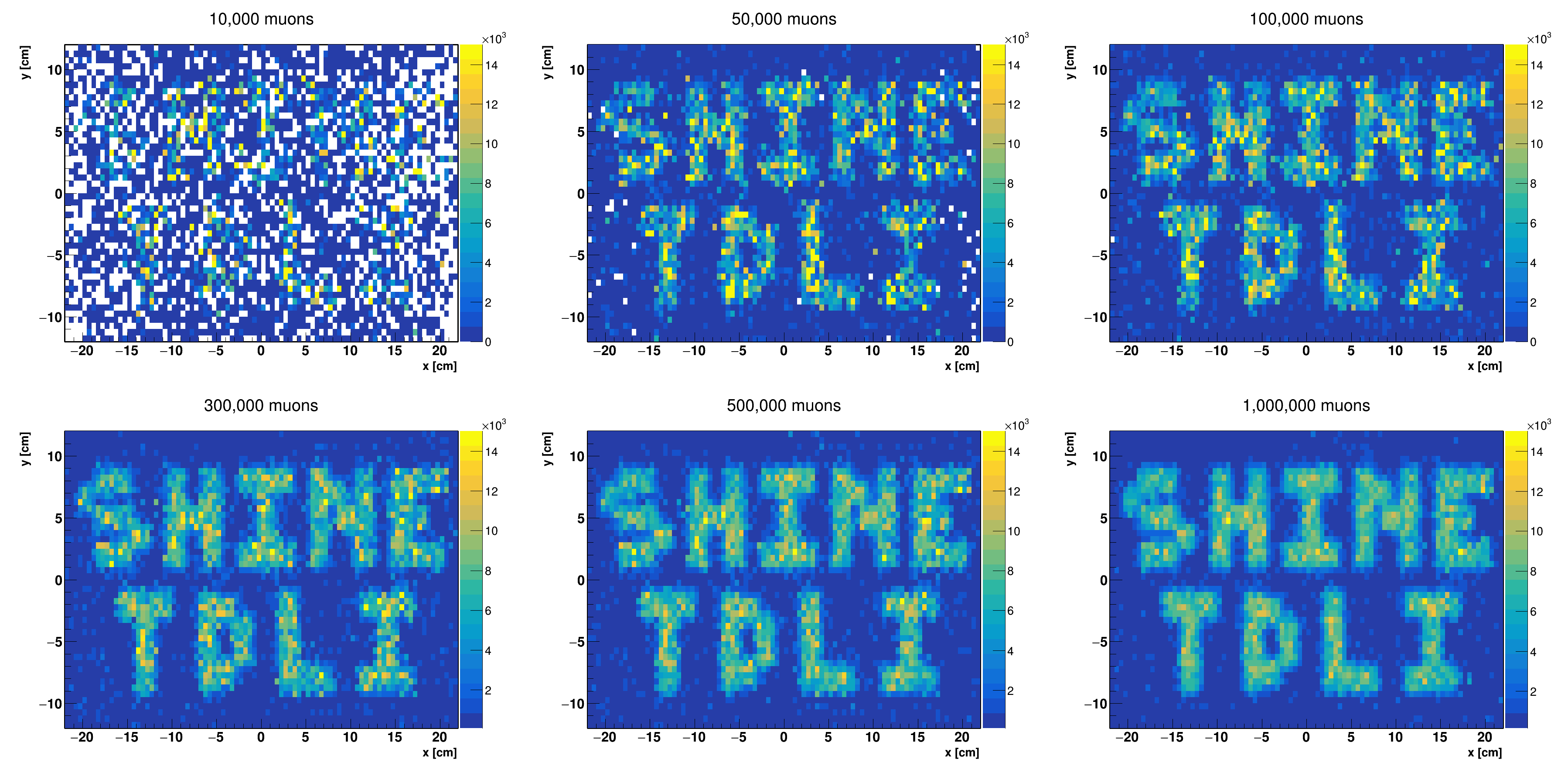}
\caption{
Scattering-tomography reconstructions of the object for increasing numbers of effective muon events. The color bar indicates the mean-square scattering angle $<\theta^{2}_{rms}>$ in the corresponding bins. The image quality improves progressively as the number of muon events increases. At $10^6$ events, the individual lead cubes are clearly resolved.
}
\label{fig:imaging_result}
\end{figure}

After the incoming and outgoing track segments are reconstructed, the Point of Closest Approach (PoCA) algorithm~\cite{Schultz2004} is used to estimate where the scattering occurred and to accumulate the corresponding scattering strength in the image volume. Figure~\ref{fig:imaging_result} shows the expected statistical evolution of the reconstruction: the object becomes progressively clearer as the number of effective muon events increases, and the individual lead cubes are clearly resolved in the $10^6$-event image.

\subsection{Imaging time estimates}
\label{sec:imaging:time}

\begin{figure}[htbp]
\centering
\includegraphics[width=0.7\linewidth]{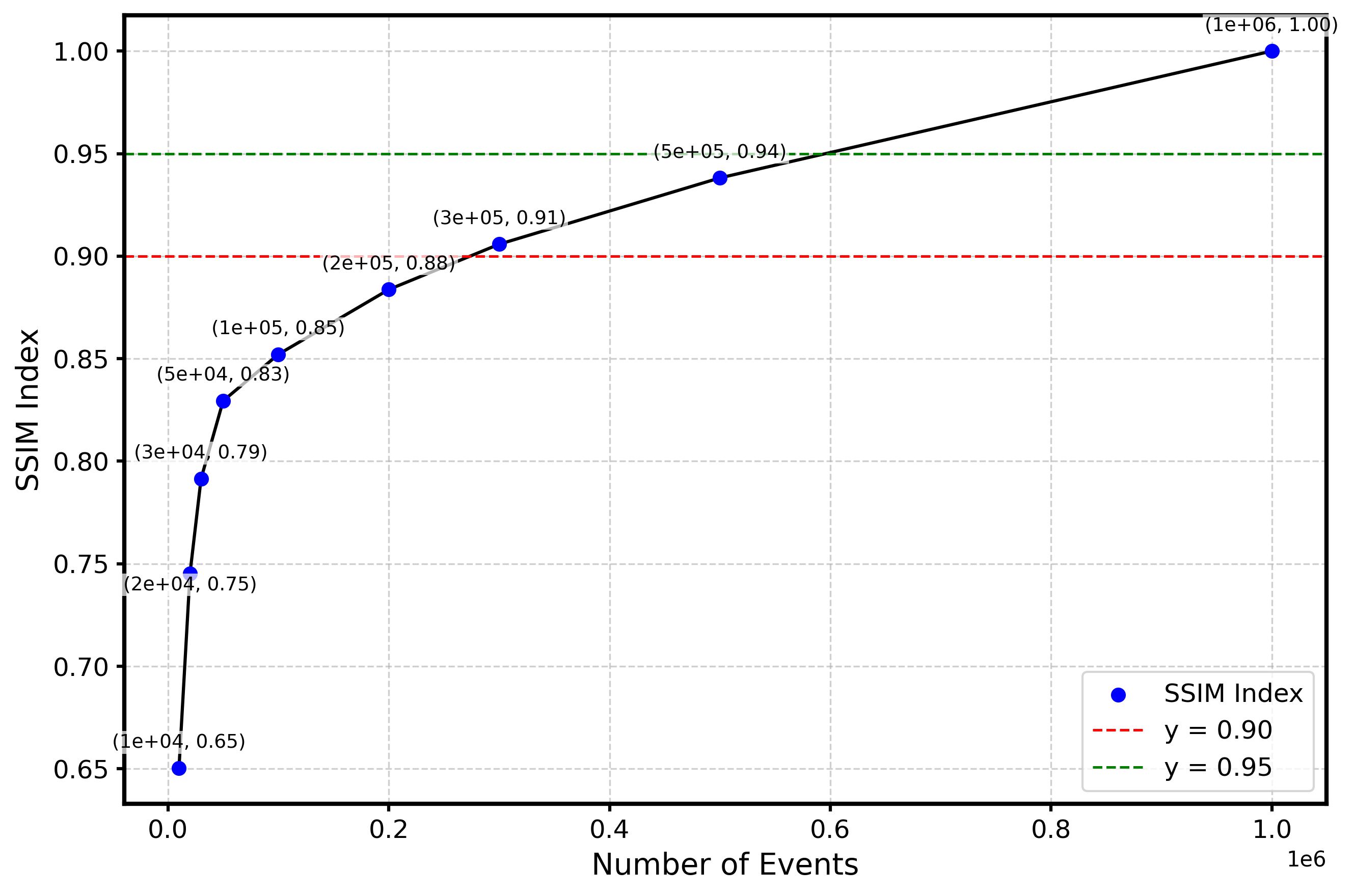}
\caption{
SSIM of the PoCA-reconstructed image as a function of the number of effective muon events, using the $10^6$-event reconstruction as the reference. The SSIM exceeds 0.9 when the number of events reaches $3\times10^5$ (dashed line), indicating satisfactory image quality.
}
\label{fig:imaging_evaluation}
\end{figure}

Using the $10^6$ muon events reconstruction as the reference, we compute the Structural Similarity Index Measure (SSIM) of the imaging results for smaller event counts. The SSIM algorithm measures the structural similarity between a distorted image and a reference image, producing a value between 0 and 1, where a higher number indicates greater similarity~\cite{Wang2004}.

Figure~\ref{fig:imaging_evaluation} shows that the SSIM exceeds 0.9 at $3	imes10^5$ effective events. We therefore use this event count as a practical criterion for a satisfactory reconstruction in the present proof-of-principle study.

Using the detectors in Fig.~\ref{fig:equipment} and the muon distribution in Fig.~\ref{fig:beam_characterization}, the effective muon rate is 0.33$\pm$0.01 muons/bunch in the bunch-by-bunch simulation. The single-muon rate is defined as the number of events containing one muon traversing all four tracker layers per electron bunch. Non-muon particles and secondary hits can contaminate the tracker and reconstruction; after these background effects are taken into account, the effective reconstructed single-muon rate is 0.28$\pm$0.01 muons/bunch. This reduction is distinct from multi-muon pileup.
To achieve a satisfactory imaging result, the \SI{3}{GeV}, \SI{50}{Hz} electron beam accumulates $3\times10^5$ tracks in approximately \SI{5.9}{\hour} (\SI{6}{\hour} allowing for experimental overhead), sufficient for a proof-of-principle demonstration of object imaging. Table~\ref{tab:imaging_time} summarizes the imaging time for the \SI{3}{GeV} commissioning configuration and an idealized design-scale lower-limit estimate for the \SI{8}{GeV}/\SI{50}{kHz} muography benchmark.

\begin{table}[htbp]
\centering
\caption{Estimated scattering-imaging times for the \SI{3}{GeV} commissioning configuration ($\sim$0.28 effective reconstructed single-muon events/bunch, corresponding to $\sim$\SI{14}{\per\second} at \SI{50}{Hz}) and an idealized design-scale lower-limit estimate for the \SI{8}{GeV}/\SI{50}{kHz} muography benchmark. The latter assumes a low-occupancy scale of approximately one muon arriving at the detector per bunch and therefore $5\times10^4$ usable-event opportunities/s; the actual acquisition time will depend on detector acceptance and reconstruction efficiency.}
\label{tab:imaging_time}
\begin{tabular}{ccc}
\toprule
$N_0$ required & Time (\SI{3}{GeV}) & Time (\SI{8}{GeV}) \\
\midrule
300\,000 & $\sim$\SI{6}{h} & $\sim$\SI{6}{s} \\
500\,000 & $\sim$\SI{10}{h} & $\sim$\SI{10}{s} \\
1\,000\,000 & $\sim$\SI{20}{h} & $\sim$\SI{20}{s} \\
\bottomrule
\end{tabular}
\end{table}

\section{Discussion}
\label{sec:discussion}

\subsection{From shielding to a usable muon beam}

One of the main outcomes of the simulation is that the \SI{3}{m} concrete wall plays a more useful role than simply separating the accelerator tunnel from the downstream experimental area. It acts as a passive particle and energy filter. The intense electromagnetic and hadronic shower produced at the target is strongly suppressed, while the more penetrating GeV-scale muons survive. This is why a downstream muography measurement becomes possible without constructing a dedicated high-energy muon transport line with bending magnets and Wien filters. In this configuration, the existing SHINE infrastructure performs part of the beam-selection task automatically.

This filtering also explains the apparently counterintuitive energy spectrum behind the wall. Muons need an incident kinetic energy of approximately \SI{1}{GeV} to penetrate the concrete, but they lose a substantial fraction of that energy while doing so. The surviving muons therefore emerge with residual kinetic energies extending from near zero to about \SI{1.2}{GeV}. The beam at the imaging location is thus the result of the combined production and filtering process, rather than a simple copy of the spectrum at the target.

\subsection{From a commissioning demonstration to a higher-rate source}

For the \SI{3}{GeV}, \SI{50}{pC}, \SI{50}{Hz} commissioning configuration, the simulated detector-arrival rate reaches approximately \SI{51}{\per\second} at the optimal position. After the detector geometry and reconstruction requirements are included, the effective reconstructed single-muon rate is about \SI{14}{\per\second}. At this rate, the $3\times10^5$ effective events required for an SSIM above 0.9 can be accumulated in approximately \SI{6}{h}, making a proof-of-principle scattering-imaging experiment feasible during commissioning.

The design-scale \SI{8}{GeV}/\SI{50}{kHz} benchmark points to a much higher available intensity, exceeding $5\times10^4~\mu/\text{s}$ under the low-occupancy assumption used here. This number should be interpreted carefully: it is a projected available source/detector-arrival intensity rather than a fully simulated effective reconstructed event rate. The idealized acquisition times listed in Table~\ref{tab:imaging_time} therefore represent lower-limit scaling estimates. Detector acceptance, reconstruction efficiency, and multi-track occupancy will ultimately determine the usable rate. Nevertheless, the comparison illustrates the substantial headroom available as SHINE moves from commissioning toward higher repetition-rate operation.

\subsection{Comparison with cosmic-ray and laser-driven sources}
\label{sec:discussion:comparison}

\begin{figure}[htbp]
\centering
\includegraphics[width=0.9\linewidth]{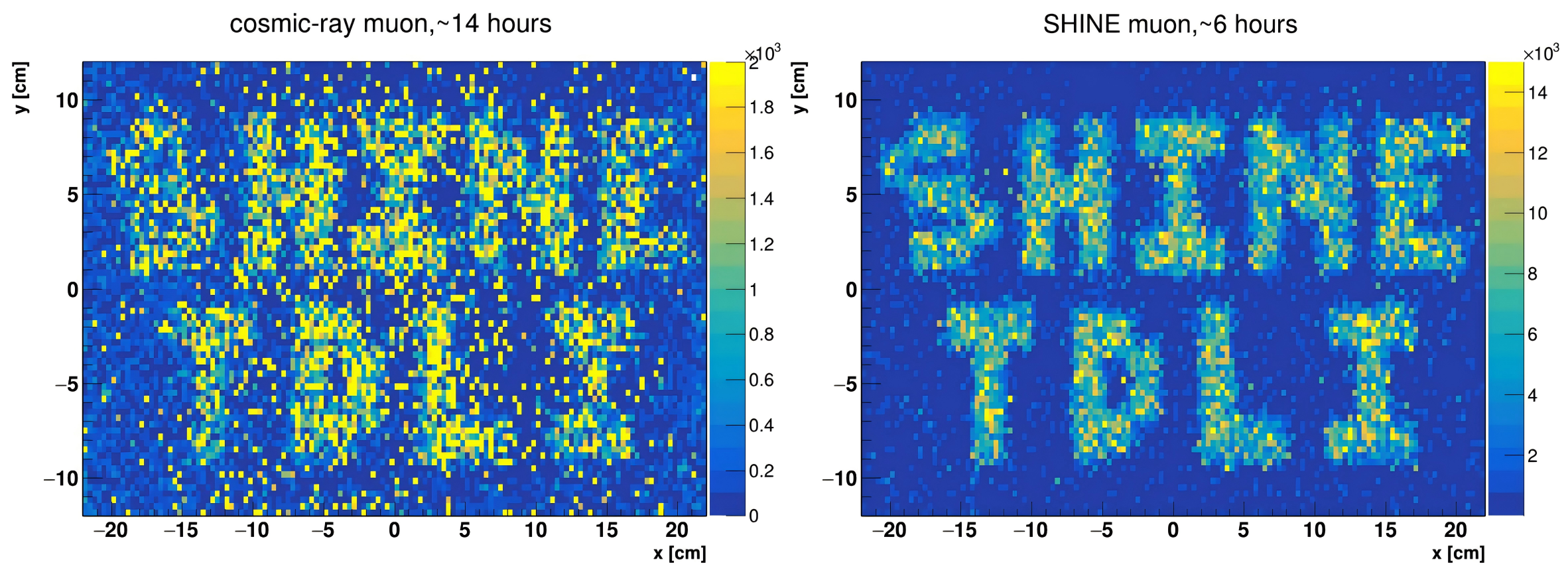}
\caption{Comparison of scattering-tomography reconstructions obtained with (a) cosmic-ray muons and (b) the simulated SHINE accelerator-based configuration, using the same number of effective muon events ($3\times10^5$). The two samples retain their respective simulated energy, angular, and spatial phase-space distributions. The SHINE case therefore represents a comparison of complete source configurations rather than a universal one-variable source-performance benchmark.}
\label{fig:CRY_SHINE}
\end{figure}

Cosmic-ray muons remain uniquely valuable because they are naturally available, highly penetrating, and require no accelerator infrastructure. Their limitations for controlled measurements are equally clear: the flux is low, the arrival direction is set by nature, and the incident phase space cannot be tuned to a particular object or detector. SHINE offers the complementary regime of a pulsed, directional, beam-synchronous source with controllable operating conditions.

For the comparison in Fig.~\ref{fig:CRY_SHINE}, cosmic-ray muons were generated with the Cosmic-ray Shower Library (CRY). The SHINE and cosmic-ray samples were not artificially matched in energy or angular distribution, and they also differ in source geometry and spatial phase space. The figure should therefore be read as a comparison between the two simulated imaging configurations. Under these conditions, the SHINE sample produces a qualitatively clearer reconstruction for the same number of effective events, while its higher available intensity can reduce the time needed to collect such a sample.

LWFA-driven muon sources provide another complementary route. Recent experiments have demonstrated muon yields of approximately $10^2$--$10^3$ muons per shot~\cite{Terzani2025PRAB,Calvin2026PPCF} and highlight the long-term possibility of compact accelerator-based muography. At present, however, variations in electron energy, charge, and pointing remain important practical limitations for systematic detector and reconstruction studies. The superconducting linac at SHINE is expected to provide substantially more reproducible pulse-to-pulse conditions, making it particularly suitable for establishing quantitative performance benchmarks.

\begin{table}[htbp]
\centering
\caption{Comparison of representative muon sources for muography.}
\label{tab:comparison}
\begin{tabular}{lccc}
\toprule
Property & Cosmic-ray & SHINE & LWFA \\
\midrule
Muon source & Atmospheric & $e^-$-on-target & $e^-$-on-target \\
Electron energy & N/A & \SIrange{3}{8}{GeV} & \SIrange{0.4}{10}{GeV} \\
Typical $\mu$ energy & $\sim$\SI{4}{GeV} & \SIrange{0}{1.2}{GeV}$^*$ & \SIrange{0.1}{8}{GeV} \\
Angular distribution & $\sim \cos^2\theta$ & Forward & Forward \\
Time structure & Continuous & Pulsed (\SI{50}{Hz}--\SI{50}{kHz}) & Pulsed (\SIrange{0.1}{10}{Hz}) \\
Beam stability & N/A & High & Limited \\
Portability & Ubiquitous & Fixed & Potentially compact \\
\bottomrule
\multicolumn{4}{l}{\small $^*$Behind the \SI{3}{m} isolation wall at \SI{3}{GeV}.}
\end{tabular}
\end{table}

\subsection{SHINE as a platform for developing accelerator muography}

The broader significance of the present study is therefore not limited to a single imaging demonstration. Electron-on-target muon production can be studied at SHINE under stable and reproducible accelerator conditions, allowing the individual parts of the technique to be optimized systematically. These include the target geometry and material, the resulting muon phase space, detector efficiency, background rejection, reconstruction algorithms, and image resolution. Such measurements can provide the quantitative baseline needed to develop accelerator muography as a technique rather than simply demonstrate that an image can be reconstructed.

The same production system also naturally spans two application regimes. The low-energy surface-muon component can be transported through the dedicated \SI{13.6}{m} beamline for $\mu$SR and other precision applications~\cite{Liu2025PRAB}, while the GeV-scale forward component considered here can be used for penetrating-muon applications. These programs therefore exploit different regions of the phase space generated by the same electron-on-target interaction.

Beyond imaging, a scattering detector of the type considered here could also support other measurements based on controlled muon deflection. The PKMu collaboration has proposed and recently demonstrated related methods with cosmic-ray muons to search for muon--dark-matter interactions through changes in the muon scattering distribution~\cite{Yu2024PKMu,Liu2026PKMu}. A high-rate, stable, and beam-synchronous source would provide a complementary environment for such studies, although this possibility lies beyond the scope of the present imaging analysis.

\subsection{Limitations and next steps}
\label{sec:discussion:limitations}

The present results are simulation-based and several sources of uncertainty should be addressed before the rate projections are treated as experimental performance. First, the predicted muon yield depends on the Geant4 physics model. The \texttt{FTFP\_BERT} physics list has been compared with FLUKA for SHINE muon-production studies, with agreement within 2\% below \SI{300}{MeV/\textit{c}} and within 30\% in the surface-muon region~\cite{Liu2025PRAB}; the corresponding uncertainty for the behind-wall high-energy component has not yet been quantified.

Second, the simulated shielding geometry uses nominal dimensions and material properties. The as-built composition of the isolation wall, including concrete density and reinforcement, can change the effective penetration threshold and therefore the transmitted muon spectrum. Third, the imaging study represents the detector response at the level currently implemented in the simulation and does not replace a full experimental validation of detector efficiencies, timing, backgrounds, and reconstruction performance.

The next step is therefore a dedicated beam test at Shaft~2. Such a measurement can validate the transmitted particle flux and energy spectrum, characterize the residual background, and test the tracking detector and reconstruction chain under realistic accelerator conditions. The present work provides the simulation framework and performance targets for that transition from source study to experimental demonstration.

\section{Summary and outlook}
\label{sec:summary}

This study follows the complete chain required for accelerator-based muography at SHINE: electron-on-target muon production, passive selection by the existing Shaft~2 shielding, characterization of the surviving beam, and scattering-based image reconstruction. The simulations show that these steps can be connected using the existing commissioning infrastructure.

At the target, photonuclear pion production provides the dominant low-energy and broadly distributed muon population, whereas Bethe--Heitler pair production contributes a more energetic and forward-directed component. The \SI{3}{m} isolation wall then strongly suppresses the accompanying electromagnetic and hadronic shower and preferentially transmits muons with incident energies above approximately \SI{1}{GeV}. In this sense, the existing Shaft~2 shielding acts as a natural muon filter.

For the \SI{3}{GeV}, \SI{50}{pC}, \SI{50}{Hz} commissioning configuration, the simulated imaging setup yields approximately 0.28 effective reconstructed single-muon events per bunch, corresponding to about \SI{14}{\per\second}. The surviving muons have residual kinetic energies extending up to \SI{1.2}{GeV}. Using this simulated phase space in a PoCA scattering-tomography reconstruction, $3\times10^5$ effective events produce an SSIM above 0.9 and can be accumulated in approximately \SI{6}{h}. This provides a realistic timescale for a proof-of-principle accelerator-muography demonstration during commissioning.

At the higher \SI{8}{GeV}/\SI{50}{kHz} benchmark, the projected available muon intensity exceeds $5\times10^4~\mu/\text{s}$ under the conservative low-occupancy assumption used in this work. The corresponding short acquisition times are idealized lower-limit estimates rather than predictions of fully reconstructed event rates, but they illustrate the potential increase in capability as the facility moves toward higher repetition-rate operation.

More broadly, SHINE offers a controlled environment in which electron-driven muography can be developed systematically. The same electron-on-target source also supports distinct low-energy and GeV-scale muon programs by exploiting different regions of the generated phase space. A dedicated beam test with plastic-scintillator tracking planes and SiPM readout is the natural next step. Such a measurement will test the predicted muon flux and backgrounds, validate the detector and reconstruction chain, and establish an experimental foundation for future accelerator-based applications in infrastructure inspection, geological imaging, and nuclear-material characterization.

\section*{Acknowledgments}

J.W., Y.W., J.K.N., J.C.Y and K.S.K.\ were supported by the Shanghai Pilot Program for Basic Research (Grant No.\ 21TQ1400221).


\bibliographystyle{JHEP}
\bibliography{biblio.bib}

\end{document}